%% file: main.tex
\documentclass[sigconf,nonacm]{acmart}
\AtBeginDocument{%
  }

\usepackage{float}
\usepackage{array}
\usepackage{longtable}
\usepackage{soul}
\usepackage{fontawesome5}
\definecolor{GriceaBlue}{HTML}{EBF4FF}
\definecolor{GriceaPurple}{HTML}{EBF4FF}
\definecolor{GriceaBlueDeep}{HTML}{77bfff}
\definecolor{GriceaBrand}{HTML}{4b0dc3}
\newif\ifgriceahighlights
\griceahighlightstrue
\newcommand{\griceaOpen}[1]{\ifgriceahighlights{\sethlcolor{GriceaBlue}\hl{#1}}\else#1\fi}
\newcommand{\griceaAccess}[1]{\ifgriceahighlights{\sethlcolor{GriceaPurple}\hl{#1}}\else#1\fi}

\newcommand{\griceaName}{\texorpdfstring{\ifgriceahighlights\textcolor{GriceaBrand}{Gricea}\else Gricea\fi}{Gricea}}
\begin{document}

\title{Gricea: An Open Science Platform for Conversational AI Research}

\author{Nikhil Sharma}
\email{nsharm27@jhu.edu}
\orcid{0009-0004-9183-6811}
\affiliation{%
  \institution{Johns Hopkins University}
  \city{Baltimore}
  \state{MD}
  \country{USA}
}

\author{Yunlin Gong}
\email{ygong37@jh.edu}
\affiliation{%
  \institution{Johns Hopkins University}
  \city{Baltimore}
  \state{MD}
  \country{USA}
}

\author{Xinyang Cheng}
\email{xcheng45@jhu.edu}
\affiliation{%
  \institution{Johns Hopkins University}
  \city{Baltimore}
  \state{MD}
  \country{USA}
}

\author{Ziang Xiao}
\email{ziang.xiao@jhu.edu}
\affiliation{%
  \institution{Johns Hopkins University}
  \city{Baltimore}
  \state{MD}
  \country{USA}
}

\renewcommand{\shortauthors}{Sharma et al.}
\begin{abstract}
We need studies on conversational AI (CAI) at scale to understand human behavior and shape CAI design. However, fragmented reporting of systems and study configurations hinders replication, extension, and knowledge accumulation. We present Gricea, an open-science platform representing studies as configurable, deployable research artifacts that researchers can run, inspect, share, and reuse. Informed by a formative analysis of prior CAI research, Gricea couples study procedures, participant-facing systems, and conversational task behavior in. In a replication study using Gricea, we replicated configurations 93\% of eligible CUI 2026 papers; while also flagging missing information in 96\% of papers that hinder faithful replication --- further motivating Gricea's need. In a user study, researchers and practitioners from diverse backgrounds successfully constructed runnable studies addressing various open-ended research questions. Together, these findings demonstrate Gricea’s support for constructing, reproducing, and extending CAI studies through shared research artifacts, enabling cumulative knowledge building through open science.
\end{abstract}


\begin{teaserfigure}
  \includegraphics[width=\textwidth]{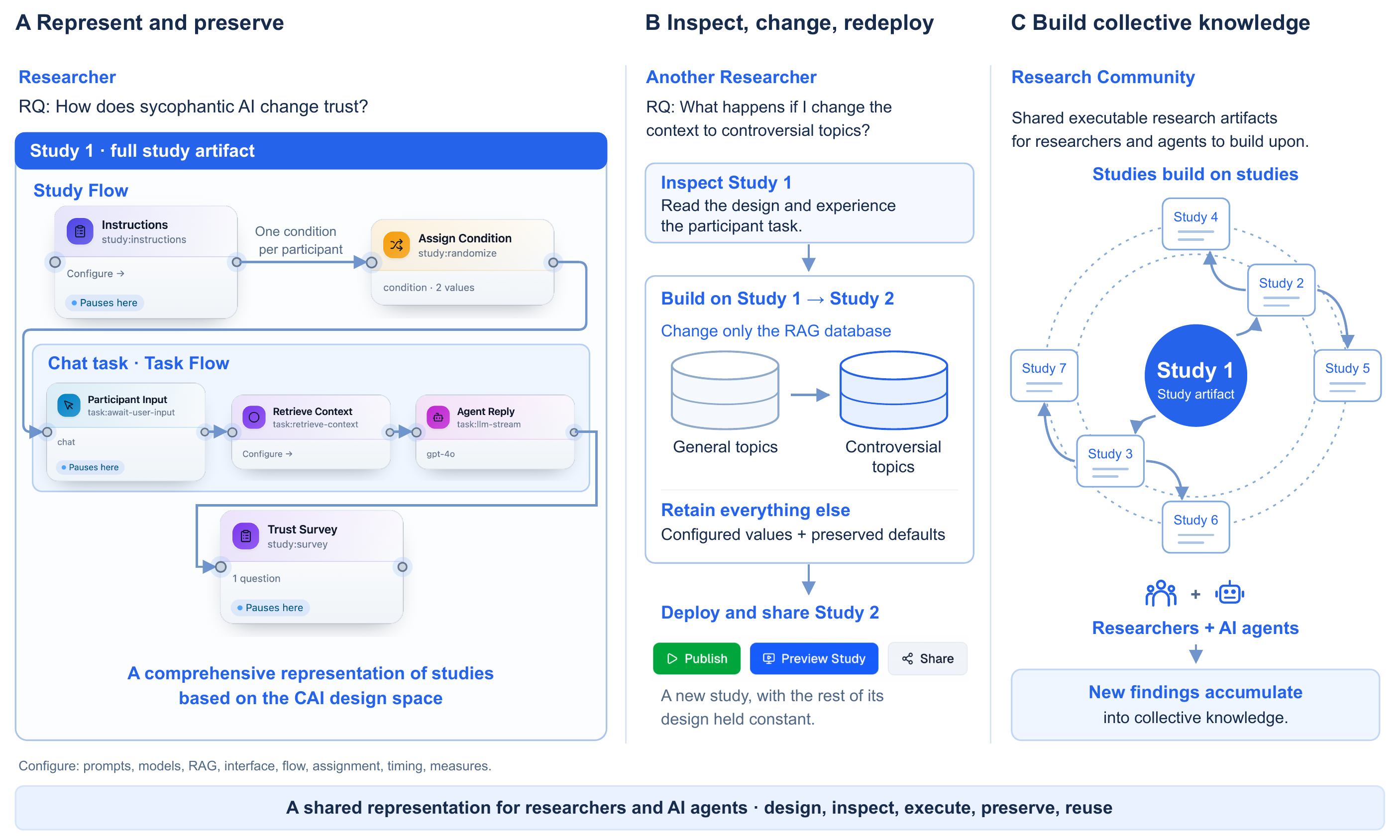}
  \caption{\textbf{A shared research foundation for conversational AI.}
  \textbf{A:} Gricea represents a study through connected Study Flow and Task Flow definitions, preserving both configured choices and unchanged defaults in an inspectable, executable artifact.
  \textbf{B:} Other researchers can inspect and extend the artifact by changing its grounding context while retaining the remaining study design; The updated study can be deployed and shared.
  \textbf{C:} The research community and AI agents can build on shared executable artifacts across successive studies, accumulating new findings into collective knowledge}
  \Description{Three panels show the lifecycle of a shared study artifact. A research configures a between-subject study using connected Study Flow and Task Flow graphs. Another researcher inspects that artifact, adds their configuration edit to the study while retaining other settings, and publishes a new version. The research community can through a shared executable representation builds collective knowledge.}
  \label{fig:teaser}
\end{teaserfigure}

\maketitle
\begin{center}
  \small
  \href{https://gricea.com}{%
    \raisebox{-0.2em}{%
      \includegraphics[height=1.2em]{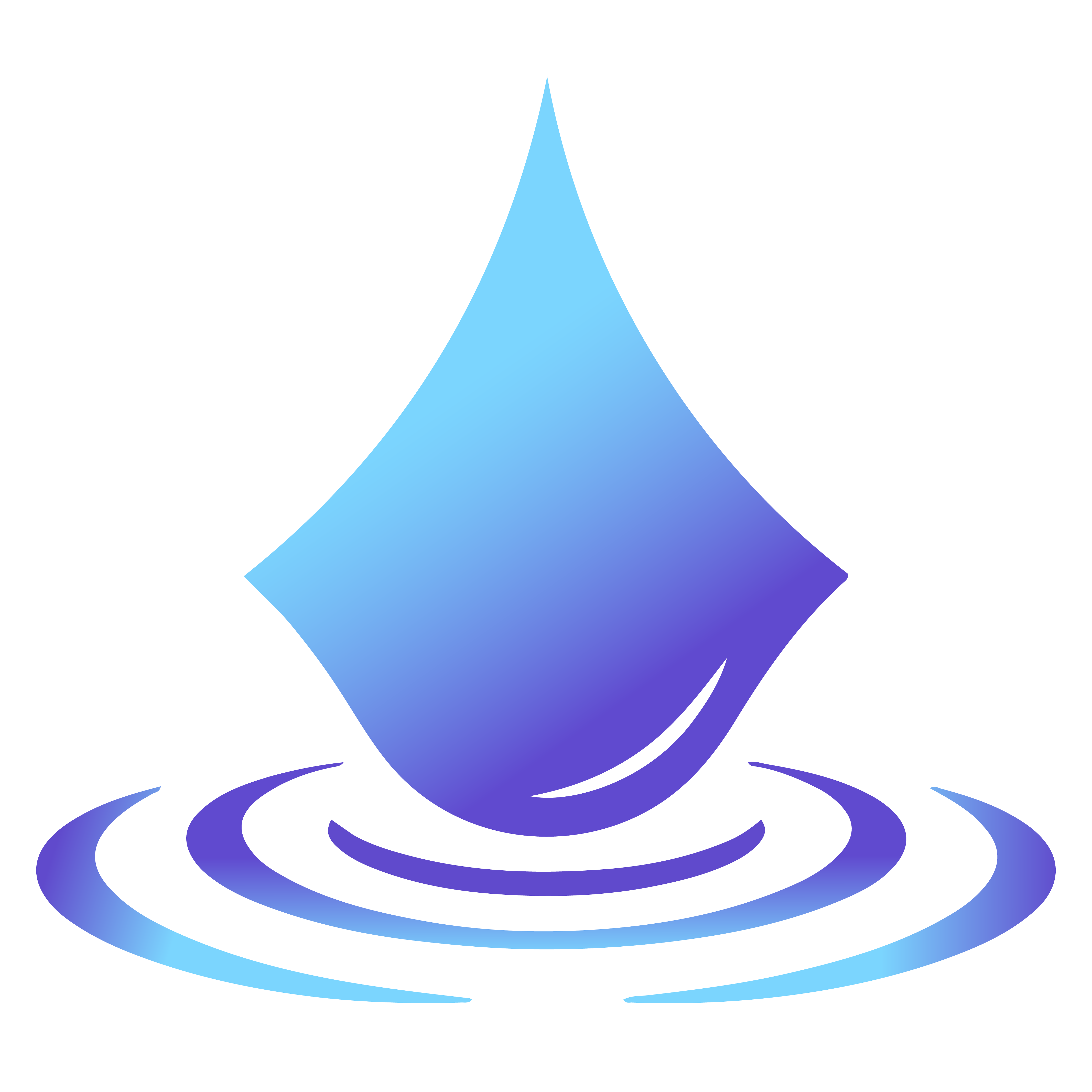}%
    }\enspace
    \textcolor{GriceaBrand}{Platform}%
  }
  \qquad
  \href{https://gricea.com/documentation/tutorial}{%
    \textcolor{GriceaBlueDeep}{\faYoutube}\enspace\textcolor{GriceaBrand}{Tutorial}}%
  
\end{center}

\input{CHI_2027/01_introduction}
\input{CHI_2027/02_related_work}
\input{CHI_2027/02_formative_study}

\input{CHI_2027/03_Gricea}
\input{CHI_2027/04_Evaluation}
\input{CHI_2027/05_discussion}
\input{CHI_2027/06_limitation}

\bibliographystyle{ACM-Reference-Format}
\bibliography{ref}

\appendix
\include{CHI_2027/appendix}

\end{document}

%% file: CHI_2027/01_introduction.tex
\section{Introduction}
Conversational AI systems built on large language models are becoming a default interface for information access, communication, and everyday work \citep{chatterji2025people}. As these systems become everyday infrastructure, they shape how people learn, create, collaborate, make decisions, and form relationships with artificial agents \citep{10.1145/3544548.3581196,kirk2025socioaffective}. Understanding these changes requires empirical research on human behavior and experience alongside evaluation of the systems people use. Such evidence is essential to the design, governance, and deployment of conversational AI.

Prior human--AI interaction research shows that outcomes depend on various factors such as model performance, how systems communicate uncertainty, present evidence, structure initiative, support verification, and distribute control \citep{Amershi2019GuidelinesFH,generative-echo-chamber,10.1145/3544548.3581388,kirk2025socioaffective}. Misinformation exposure, overreliance, persuasion, and uneven information access emerge through the interplay of model behavior, interface design, and user context \citep{information-diet,faux-polyglot,sharma2026feedback,zhang2025personalizationlargelanguagemodels,shi2025sirensongllmsusers}. Studying these effects requires examining how people interact with conversational AI while pursuing goals in specific tasks and contexts; researchers use Controlled human-subject studies to isolate effects of different design choices \citep{liao2025rethinkingmodelevaluationnarrowing}.

Building cumulative knowledge using controlled studies about conversational AI requires a shared frame of reference for what participants experienced, how conditions were configured, and how outcomes were measured. When those configurations and materials are difficult to recover, subsequent researchers must reconstruct the interaction before they can reproduce or extend a study. Differences in interface behavior, agent configuration, or procedure can then obscure what was preserved and what changed. Open science therefore requires preserving these methodological choices in an inspectable, reusable form alongside the findings \citep{foster2017osf,Aguilar_2024}. A shared representation gives researchers a common starting point for reproducing conditions, systematically varying design choices, and relating new findings to prior work.

However, representing these conditions is challenging because design choices span several connected parts of a study: the interface and actions available to participants, agent behavior context, task, modality, participant population, and experimental procedure. Furthermore, all of these conditions can interplay with each other creating a vast and diverse design space. A shared representation must make the design choices both at the study level and at the task configuration level explicit; preserving a comprehensive representation of based on the CAI design space (Section ~\ref{sec:formative}).

Even without the overhead of a shared representation, conducting these studies has high costs beyond just recruiting and compensating participants. Researchers must coordinate study procedures, task interfaces, agent behavior, and data collection, often by integrating survey tools, custom interfaces, model services, and deployment infrastructure. Commercial conversational systems offer limited experimental control, while changes to their interfaces, retrieval policies, or models can alter the conditions under investigation. Building and maintaining custom systems therefore demands time and engineering expertise that can constrain both who conducts conversational AI research and which questions they pursue. Therefore, a successful shared representation artifact must not only find a scalable way to represent the vast and diverse design space but also reduce the costs of building these custom systems while making the artifact a by-product of the process rather than an additional cost to the researchers.


In this paper we present \textit{\griceaOpen{Gricea}}, a platform that represents conversational AI study designs as configurable, deployable research artifacts that researchers can run, inspect, share, and reuse. Researchers visually author study procedures and configure complex interactive task conditions within a platform that supports participant-facing execution and instrumentation. The representation researchers inspect is also the definition the runtime executes, connecting experimental assignment and task behavior to the participant experience. \griceaAccess{Gricea} reduces the technical overhead of constructing controlled studies and preserves immutable study versions as reusable templates. Other researchers can inspect the original design, reproduce its conditions, and extend it to new questions, making individual studies reusable resources enabling cumulative knowledge (Figure~\ref{fig:teaser}).

\griceaOpen{Gricea's} design was informed by a formative analysis of 57 papers on conversational AI (Section~\ref{sec:formative}). The resulting desiderata guided two complementary representations. \textit{Study Flow} represents the experimental procedure, connecting surveys, tasks, and between- and within-subject structures. \textit{Task Flow} represents behavior within each task, including agent configuration and the participant-facing interface. Researchers can combine these elements to implement existing designs and complex configurations involving multiple agents, multiple users, and customized interfaces.

We evaluated \griceaAccess{Gricea} with $N=10$ researchers and practitioners from diverse disciplinary backgrounds, who completed an assisted authoring walkthrough before independently designing studies around their own research questions. Participants implemented valid, runnable studies that varied across procedures, interfaces, models, surveys, and outcomes, and the no-code interface reduced barriers for researchers without systems backgrounds. We also examined CUI 2026 full papers, none of which informed \griceaOpen{Gricea's} design. Of 29 eligible papers, we replicated designs from 27 as executable artifacts: 10 completely and 17 partially. Together, these evaluations demonstrate support for authoring diverse studies and reconstructing published designs for inspection and reuse.

This paper makes three contributions. \textbf{First}, we contribute a shared, executable representation of conversational AI studies that connects study procedure, task behavior, participant-facing conditions, and instrumentation, enabling researchers to inspect, reproduce, and extend experimental designs. \textbf{Second}, we present \griceaAccess{Gricea}, a configurable no-code research platform that operationalizes this representation through visual authoring, participant-facing execution, immutable study versions, shareable configurations, and reusable community templates. These mechanisms lower technical barriers while making individual studies available as resources for subsequent research. \textbf{Third}, we contribute empirical findings from an authoring study with researchers and practitioners from diverse disciplinary backgrounds and reproductions of CUI 2026 study configurations, demonstrating \griceaOpen{Gricea’s} support for diverse research questions and published study designs while identifying remaining needs for guidance, validation, and workflow support.

%% file: CHI_2027/02_related_work.tex
\section{Related Work}
\label{sec:related-work}

\griceaAccess{Gricea} builds on research that broadens participation in online studies, preserves methods as reusable research resources, and makes conversational systems configurable through higher-level representations. These efforts address complementary requirements for conducting research and building on its findings. \griceaOpen{Gricea} brings these requirements together through a shared representation of study procedure and conversational task behavior. The representation used to design a study governs its deployment and data collection, preserving an inspectable research artifact as part of building and running the study.

\subsection{Infrastructure for Crowdsourced and Online Studies}

Online research infrastructure has expanded where studies can be conducted and who can participate. \citet{kittur2008crowdsourcing} examined how task design and quality checks influence crowdsourced judgments, while \citet{reinecke2015labinthewild} used personalized feedback to attract uncompensated participants and evaluated online replications of laboratory studies. Subsequent comparisons examined differences in participant diversity and data quality across recruitment platforms \citep{peer2017beyond,peer2021quality}. These efforts demonstrate the influence of recruitment in the quality of the online studies.

Conducting experiments with these participant populations also requires infrastructure for implementing tasks, assigning conditions, coordinating interactions, and collecting responses. Reusable experiment frameworks address these requirements by providing components that researchers can adapt across studies. oTree supports browser-based experiments through Python and HTML, including a library of reusable game templates \citep{Chen_Schonger_Wickens_2016}. Empirica supports configurable experimental designs and reusable protocols for real-time group experiments \citep{Almaatouq_Becker_Houghton_Paton_Watts_Whiting_2021}, while jsPsych enables researchers to construct behavioral experiments from reusable plugins and contribute new tasks to a community ecosystem \citep{de_Leeuw_2023}. Across these systems, reusable components allow the implementation work behind one study to support subsequent studies, reducing the effort required to develop and extend experimental designs.

Human--AI research platforms bring agent behavior into this experimental infrastructure. Deliberate Lab combines no-code experimental stages, human and LLM participants, agent mediators, and cohort management for studying human--AI group dynamics \citep{Qian_Tsai_Behr_Hussein_Laugier_Thain_Dixon_2025}. For conversational AI studies, the experimental condition depends on more than the sequence of study stages or an agent's configuration. To represent a broad set of CAI studies, we conduct a formative study to uncover the design space, allowing \griceaAccess{Gricea} to extend existing efforts to a more general reusable research infrastructure through a coupled representation of study procedure and conversational task behavior. 

\subsection{Infrastructure for Open Science}

Open-science infrastructure supports the preservation and exchange of research materials across teams. \citet{foster2017osf} describe infrastructure for project organization, collaboration, file versioning, and registration, making materials easier to preserve and share. However, accessible materials must also be sufficiently specified and connected for others to use them. \citet{reprohci2026} identified obstacles to computational reproduction among CHI papers that shared data and analysis code, illustrating the difference between making resources available and enabling others to reproduce the work. For conversational AI studies, researchers need to understand how the procedure, interface, agent behavior, and materials jointly determined what participants experienced.

Executable research representations connect methodological specification to implementation. \citet{Aguilar_2024} represent experiment components through automation code and digital documentation, including infrastructure, data collection, analysis, and management. \citet{Nobre_2021} support inspecting participant behavior through interaction provenance and replay, while \citet{Cutler_2026} connect study specification, execution, analysis, and dissemination within a browser-based framework. Subsequent LLM integration preserves conversation history and supports replay of chatbot interactions \citep{he2026revisitllm}. \griceaOpen{Gricea} builds on this connection between executable methods and inspectable interactions through a shared representation of conversational task logic and the surrounding experimental procedure which is also the same representation that the runtime executes.

Reporting frameworks and agent-native research artifacts further clarify what must survive publication. \citet{guidellm2026} call for explicit documentation of LLM use, including model versions, prompts, and configurations, while \citet{liu2026agentnative} connect scientific logic, executable code, exploration traces, and evidence so that humans and AI agents can understand and build on research. \griceaAccess{Gricea} integrates artifact preservation into the development and execution of participant studies. The configured procedure, prompts, materials, and interaction logic constitute the study that is deployed, so researchers do not need to reconstruct a separate artifact after implementation. Sharing that representation makes the implemented method available for inspection, reconfiguration, and reuse within the same environment.

\subsection{Infrastructure for visual programming of CAI studies}

Visual and declarative systems make computational choices accessible through representations that users can inspect and modify. \citet{wu2022promptchainer} support composing and debugging multi-step LLM chains, while \citet{10.1145/3613904.3642016} support systematic comparison of prompt and model variations through a visual dataflow environment. \citet{cai2024lowcodellmgraphicaluser} allow users to edit a proposed workflow before an LLM executes it, and \citet{feng2025canvildesignerlyadaptationllmpowered} support structured specification and testing of model behavior within interface design work. Conversational application platforms also provide deployment environments, and live-traffic experiments \citep{google_dialogflow_playbooks,google_dialogflow_versions,google_dialogflow_experiments}. \griceaOpen{Gricea} brings this control over computational behavior into the representation of a human-subject study, connecting experimental assignment, participant interaction, and measurement.

Research-oriented representations bring methodological choices into these abstractions. \citet{jun2019tea} allow researchers to declare study designs, assumptions, and hypotheses for statistical analysis. \citet{yao2026throughlens} provide an experiment configuration language and controls over collaborative environments, agent perception and action, and synchronized interaction logs, while \citet{zhang2024crew} support configurable human--AI teaming environments and feedback collection. \griceaAccess{Gricea} separates and couples study procedure and conversational task logic through the same graphs that drive execution. Researchers can examine how a procedural decision changes the participant-facing condition and preserve that relationship when a study is shared, reproduced, or extended. The formative analysis that follows identifies the recurring study requirements that informed this design.

%% file: CHI_2027/02_formative_study.tex
\section{Formative Analysis: The Science of Conversational AI Studies}
\label{sec:formative}
To scope the infrastructural requirements for \griceaAccess{Gricea}, we conducted a formative design space analysis of papers on conversational AI systems. Our goal was to identify recurring patterns across prior work: what studies on conversational AI investigate, what they manipulate, how they are typically conducted, what technical demands those choices create, and what forms of infrastructure existing systems already provide. From these recurring patterns, we identified what researchers need to specify and control, and which details must remain inspectable for others to reproduce and build on a study. These requirements motivate five design desiderata for \griceaOpen{Gricea’s} study representation and authoring environment.

\paragraph{Analysis Procedure.}
We began by collecting 100 candidate papers using keyword combinations around \textit{conversational AI}, \textit{agent}, or \textit{LLM}, together with terms related to \textit{users}, \textit{humans}, and \textit{studies}, across venues and repositories such as CHI, UIST, CUI. We then filtered this set to 57 papers that centered participant-facing conversational or agentic AI systems and provided sufficient detail about the study design, system configuration, or evaluated interaction condition. For each paper, the research team coded the study type, focus area, participant count, independent and dependent variables, between- and within-subject structure, procedural stages, system or pipeline components, analysis methods, and the overall structure of the study procedure. The research team reviewed and clustered these codes to identify recurring outcome areas, manipulation dimensions, procedural structures, and infrastructural demands.

As part of this analysis, we also examined how papers visually represented their study designs. Papers used staged diagrams, branching structures, and flowcharts to communicate condition assignment, task sequences, and follow-up measures. These representations make explicit how study logic structures the activities participants experience; motivating \griceaAccess{Gricea’s} support for executable visual representations: researchers should be able to design, inspect, communicate, and run a study through the same representation, without reconstructing its logic manually in code.

\subsection{What studies on conversational AI investigate}

Studies on conversational AI investigate how configured assistants shape human behavior, judgment, and experience within particular task settings. In our corpus, these settings included co-writing, conversational search, learning, dietary recommendation, and daily planning and reflection. Participants composed text with generated suggestions\citep{10.1145/3544548.3581196}, explored information through dialogue, received personalized recommendations\citep{10.1145/3719160.3736635}, or revisited plans across sessions\citep{10.1145/3719160.3736634}. Each task establishes what participants are trying to accomplish and the role the assistant plays in that activity.

Within these settings, the outcomes of interest are similarly broad. Prior work examines trust, reliance, persuasion, misinformation response, privacy behavior, writing quality, learning, and collaboration \citep{10.1145/3544548.3581196,generative-echo-chamber,faux-polyglot,sharma2026feedback,zhang2025personalizationlargelanguagemodels,shi2025sirensongllmsusers}. What links these studies is not a single application domain, but a common methodological concern: how a conversational system condition shapes what users believe, do, and produce over time. A platform for this area must therefore support both configuring the participant-facing interaction and collecting the evidence needed to examine its outcomes, including self-reports, behavioral traces, and task outputs \citep{10.1145/3719160.3736623,liu2026behavioral,10.1145/3613904.3642625}.

\subsection{The manipulation space of conversational AI studies}

Our formative analysis shows that studies on conversational AI manipulate far more than prompts or underlying models. The true experimental object is a \textit{configured interaction condition}: the combination of agent behavior, interface, context, and procedure that defines what participants experience. For example, a study of chatbot relationship framing varied both the agent's self-description and the visibility of conversation history across sessions \citep{10.1145/3719160.3736617}. Therefore, representation of studies requires specifying both what researchers manipulate and the surrounding configuration they hold constant.

We synthesize recurring configurations into six interacting dimensions: \textit{Interface Condition}, \textit{Agent Condition}, \textit{Context \& Grounding}, \textit{Task \& Modality}, \textit{Domain \& Audience}, and \textit{Study Procedure}. Table~\ref{tab:cai-study-space} summarizes their configurations and infrastructural implications. These dimensions connect what researchers configure, what participants experience, and how the study is conducted.

\begin{table*}[t]
\centering
\small
\renewcommand{\arraystretch}{1.15}
\begin{tabular}{p{\dimexpr.19\linewidth-2\tabcolsep\relax} p{\dimexpr.40\linewidth-2\tabcolsep\relax} p{\dimexpr.41\linewidth-2\tabcolsep\relax}}
\toprule
\textbf{Dimension} &
\textbf{Examples of study configuration} &
\textbf{Implications for study infrastructure} \\
\midrule

\textbf{Interface Condition} &
Layout and workspace organization; citation and provenance display; highlighting; progress guides; verification prompts; structural overlays; editability of shared artifacts; time-synchronized guidance and conversation navigation \citep{chen2025generativeinterfaceslanguagemodels,biscuit-scaffolding-llm,10.1145/3706598.3714222,10.1145/3719160.3736624,10.1145/3719160.3737636}. &
Represent what participants can see, inspect, and change. Preserve presentation and available actions as part of the study condition, including how evidence, guidance, and task progress are displayed. \\
\midrule

\textbf{Agent Condition} &
System instructions; persona or role framing; stance; response style; initiative; clarification behavior; trust-adaptive interventions; tool-use policy; response delays and waiting cues; coordination of model components \citep{10.1145/3544548.3581196,10.1145/3613904.3642134,li2025singleturnsurveymultiturninteractions,10.1145/3706598.3713760,10.1145/3719160.3736636,10.1145/3719160.3737635}. &
Make agent behavior, permitted actions, timing, and coordination explicit. Distinguish the roles participants encounter from the model components that implement those roles. \\
\midrule

\textbf{Context \& Grounding} &
Retrieved documents; retrieval policy; context filtering; citation constraints; external evidence injection; opposing or balanced context; task context; interaction history; persistent memory and its inspection or editing \citep{lewis2020retrieval,generative-echo-chamber,faux-polyglot,10.1145/3719160.3736624,10.1145/3719160.3737617}. &
Specify what information the agent receives, what participants can inspect or change, and how context persists across turns or sessions. Distinguish information available to the agent from history displayed to participants and records retained for research. \\
\midrule

\textbf{Task \& Modality} &
Search, writing, tutoring, coding, annotation, voice, multimodal interaction, direct artifact manipulation, and conversational control of haptic feedback \citep{Qian_Tsai_Behr_Hussein_Laugier_Thain_Dixon_2025,yao2026throughlens,Flores_Saviaga_2025,caetano2025agenticworkflowsconversationalhumanai,10.1145/3719160.3736638}. &
Connect different participant activities and modality-specific inputs and outputs to a shared study procedure. Support task-specific actions and outcome collection without rebuilding the surrounding study infrastructure. \\
\midrule

\textbf{Domain \& Audience} &
Topic domain; expertise; prior beliefs and experience; participant population; targeted community; high-stakes versus low-stakes settings; language and dialect; accessibility and care needs \citep{Pang_Schroeder_Smith_Barocas_Xiao_Tseng_Bragg_2025,information-diet,Ge_2024,Flores_Saviaga_2025,10.1145/3719160.3736615,10.1145/3719160.3736605}. &
Document the task setting, study materials, and relevant participant characteristics separately from assigned conditions. Preserve the context needed to interpret findings and adapt a study for another population or setting. \\
\midrule

\textbf{Study Procedure} &
Pre-task elicitation; condition assignment; branching; within- or between-subject structure; counterbalancing; post-task measures; longitudinal follow-up; scheduled check-ins and cross-session continuity \citep{10.1145/3544548.3581196,Qian_Tsai_Behr_Hussein_Laugier_Thain_Dixon_2025,10.1145/3544548.3581427,10.1145/3719160.3736634,10.1145/3719160.3736617,10.1145/3719160.3736625}. &
Make assignment, task order, branching, measurement timing, and cross-session continuity explicit and authorable. Preserve the procedure as executable logic that can be inspected, reused, and extended. \\
\bottomrule
\end{tabular}
\caption{Recurring dimensions of conversational AI study configuration, synthesized from the formative analysis. Examples include experimental manipulations, fixed settings, and participant characteristics. The infrastructure implications identify what must remain explicit to design, inspect, reproduce, and extend a study.}
\label{tab:cai-study-space}
\end{table*}

Across these dimensions, understanding a study requires inspecting how its procedure, interface, agent behavior, and contextual information jointly produce the participant experience \citep{10.1145/3544548.3581196,generative-echo-chamber,10.1145/3719160.3736617}. Researchers need to distinguish the choices that define a condition from those held constant, and to understand how those choices are implemented. A shared study representation should preserve these relationships so that both the original research team and subsequent researchers can inspect the design, reproduce its conditions, and make deliberate changes when extending it.

\subsection{The Procedural Anatomy of Conversational AI Studies}

The analysis also shows that studies on conversational AI combine system configurations with structured research procedures. A study may begin with consent, instructions, and pre-task elicitation, proceed through condition assignment and an interactive task, and conclude with post-task measures, reflection, or interviews. Prior work combines these elements in co-writing tasks with pre- and post-task measures \citep{10.1145/3544548.3581196,10.1145/3613904.3642134}, conversational search with turn-level behavior and post-task attitudes \citep{generative-echo-chamber,10.1145/3512913}, and daily planning and reflection that link repeated conversations to daily surveys and an exit interview \citep{10.1145/3719160.3736634}. A study on conversational AI therefore couples a \textit{study procedure} with an \textit{interactive runtime}.

The study procedure determines how participants move through the study: the instructions they receive, the condition they encounter, when branching occurs, and when measurements are collected. The interactive runtime determines what participants experience within a task: what they can see and do, what information the system receives, how it responds, how the interface adapts, and which actions and responses are logged. Together, these levels specify both how participants encounter a condition and how that condition operates during the interaction.

Across these studies, condition assignment and information collected before the task can configure the agent's behavior and available context, while participants' actions within the task can determine subsequent interaction paths and when post-task measures are collected \citep{10.1145/3706598.3714222,10.1145/3719160.3736635,10.1145/3719160.3736634}. A research platform must therefore represent these dependencies explicitly, connecting the assigned condition, the interaction participants experience, and the evidence collected about its outcomes. This requirement motivates \griceaOpen{Gricea's} coupled support for \textit{Study Flow} and \textit{Task Flow}, allowing researchers to specify the surrounding procedure and within-task behavior as connected parts of the same study.

\subsection{Synthesis of Infrastructural Challenges}

Executing studies across this broad manipulation space is challenging because the experimental condition is distributed across many components that must be built and controlled in tandem. Researchers often need to stitch together survey tools, custom interfaces, backend orchestration, model and retrieval pipelines, assignment logic, deployment infrastructure, and fine-grained behavioral logging \citep{Qian_Tsai_Behr_Hussein_Laugier_Thain_Dixon_2025,yao2026throughlens}. Even when the intended manipulation is conceptually straightforward, implementing it as a controlled, reproducible participant experience requires substantial engineering effort, creating barriers for researchers without the technical expertise or resources to build and maintain this infrastructure.

Maintaining experimental control also requires researchers to specify how interface affordances, retrieval behavior, model configuration, and system defaults jointly produce the participant experience \citep{Qian_Tsai_Behr_Hussein_Laugier_Thain_Dixon_2025,yao2026throughlens}. Reliance on commercial systems adds dependencies whose behavior may not be fully exposed or preserved across versions, with documented changes in model behavior showing why the same prompt and model name do not establish an equivalent condition \citep{Chen_2024}. Study infrastructure must therefore make the configuration and its dependencies inspectable alongside the record of what participants actually encountered.

When instructions, prompts, task materials, interface behavior, and procedural logic remain embedded in one-off implementations, publishing the findings does not necessarily preserve the condition needed to reproduce or extend the study. Subsequent researchers must reconstruct how these elements were connected before they can determine whether a new implementation reproduces the original condition or introduces consequential differences. Preserving the configured study as an inspectable, executable artifact provides a shared frame of reference for comparing implementations, adapting procedures, and building on prior work \citep{Qian_Tsai_Behr_Hussein_Laugier_Thain_Dixon_2025,yao2026throughlens}. Hence, such infrastructures must reduce the effort of constructing studies along with preserving the configuration necessary for collective knowledge to accumulate.

%% file: CHI_2027/03_Gricea.tex
\section{Design Desiderata}

Building on the formative analysis (Section~\ref{sec:formative}), we derive five design desiderata for infrastructure that supports the design, execution, and reuse of conversational AI studies.

\subsection*{D1: Explicit Representation of Study Conditions.}

The platform must represent the configured interaction condition shown to participants rather than only isolated prompts, screens, or model calls. Prior studies manipulate agent behavior and participant-facing interaction support while also specifying shared interfaces, controls, and contextual information across conditions \citep{10.1145/3544548.3581196,10.1145/3719160.3736624}. The infrastructure must therefore make the relationships among procedure, interface, agent behavior, and context explicit, allowing researchers to distinguish what is manipulated from what is held constant. Researchers must be able to inspect how these choices shape what participants see, what actions they can take, and how the system responds, without reconstructing the condition from separate implementation details. This representation must support checking whether the implemented assignment, task behavior, and measures correspond to the intended experimental design before publication.

\subsection*{D2: Coupled Support for Study Flow and Task Flow.}

The platform must support both the overall study procedure and the behavior of the interactive task, while keeping their roles distinct. \textit{Study Flow} specifies how participants move through instructions, condition assignment, tasks, and measures, whereas \textit{Task Flow} specifies how participant actions, agent responses, and branching shape progression within a task. Prior studies connect information collected before an interaction to the agent's behavior and assess the resulting experience through subsequent measures \citep{10.1145/3719160.3736635}. Researchers must therefore be able to author and modify each flow separately while specifying how information enters a task, when the task ends, and how its outputs connect to subsequent study stages.

\subsection*{D3: Lower Technical Barriers to Controlled Study Authoring.}

The platform must reduce the amount of bespoke engineering required to build and deploy controlled studies on conversational AI \citep{Qian_Tsai_Behr_Hussein_Laugier_Thain_Dixon_2025}. Our formative analysis identified staged diagrams, branches, and condition flows as recurring ways of representing study logic. The authoring model should build on these representations, enabling researchers without systems backgrounds to specify and connect study components through a no-code interface. Reusable support for participant interfaces, model integration, deployment, and data collection should allow researchers to move from a study concept to an executable artifact without reconstructing the surrounding infrastructure from scratch; Lowering the barriers for conducting these studies allowing for a broader range of researchers to contribute.

\subsection*{D4: Reproducibility Through Inspectable and Reusable Artifacts.}

The platform must preserve authored studies as explicit research artifacts rather than leaving critical details embedded in transient setup steps or one-off implementations \citep{Cutler_2026}. Each study version must retain its procedure, task logic, prompts, model and context settings, and participant-facing materials in an executable form. Recorded interactions must remain linked to the corresponding version so that researchers can inspect both the authored condition and the experience participants encountered. Another researcher should be able to inspect, redeploy, adapt, and build on the artifact, with changes made explicit across versions. Preserving this continuity provides a shared frame of reference for reproducing studies and building cumulative knowledge.

\subsection*{D5: Extensibility Across Tasks, Modalities, and Study Settings.}

The platform must remain extensible as the design space of conversational AI continues to expand. Studies may involve text chat, coding, multimodal interaction, voice, longer-running procedures, or community-facing workflows. New task types, modalities, and study settings should be supported through extensions that reuse the platform's mechanisms for study execution and data collection, rather than requiring reimplementation of the surrounding system \citep{yao2026throughlens}. These extensions must remain configurable within the study representation, allowing researchers to inspect, version, and reuse the resulting studies through the same authoring environment.
\begin{figure*}[t]
    \centering
    \includegraphics[width=\linewidth]{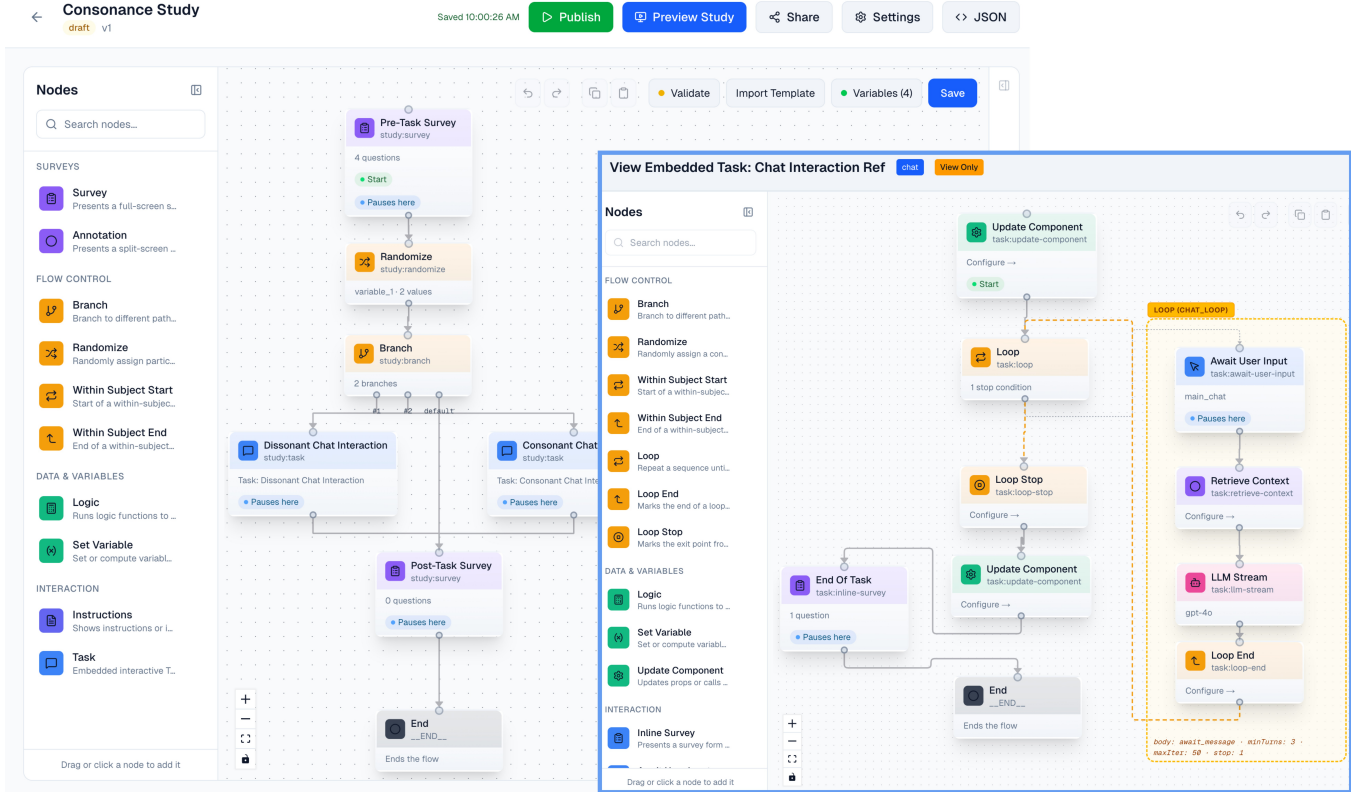}
    \caption{The \griceaAccess{Gricea} authoring interface, illustrating how researchers can assemble multi-stage study procedures (left) and independently configure agent pipelines and interface scaffolds for specific conversational tasks (right).}
    \Description{The Gricea authoring interface, illustrating how researchers can assemble multi-stage study procedures (left) and independently configure agent pipelines and interface scaffolds for specific conversational tasks (right).}
    \label{fig:study-authoring}
\end{figure*}

\begin{figure*}[h]
    \centering
    \includegraphics[width=\linewidth]{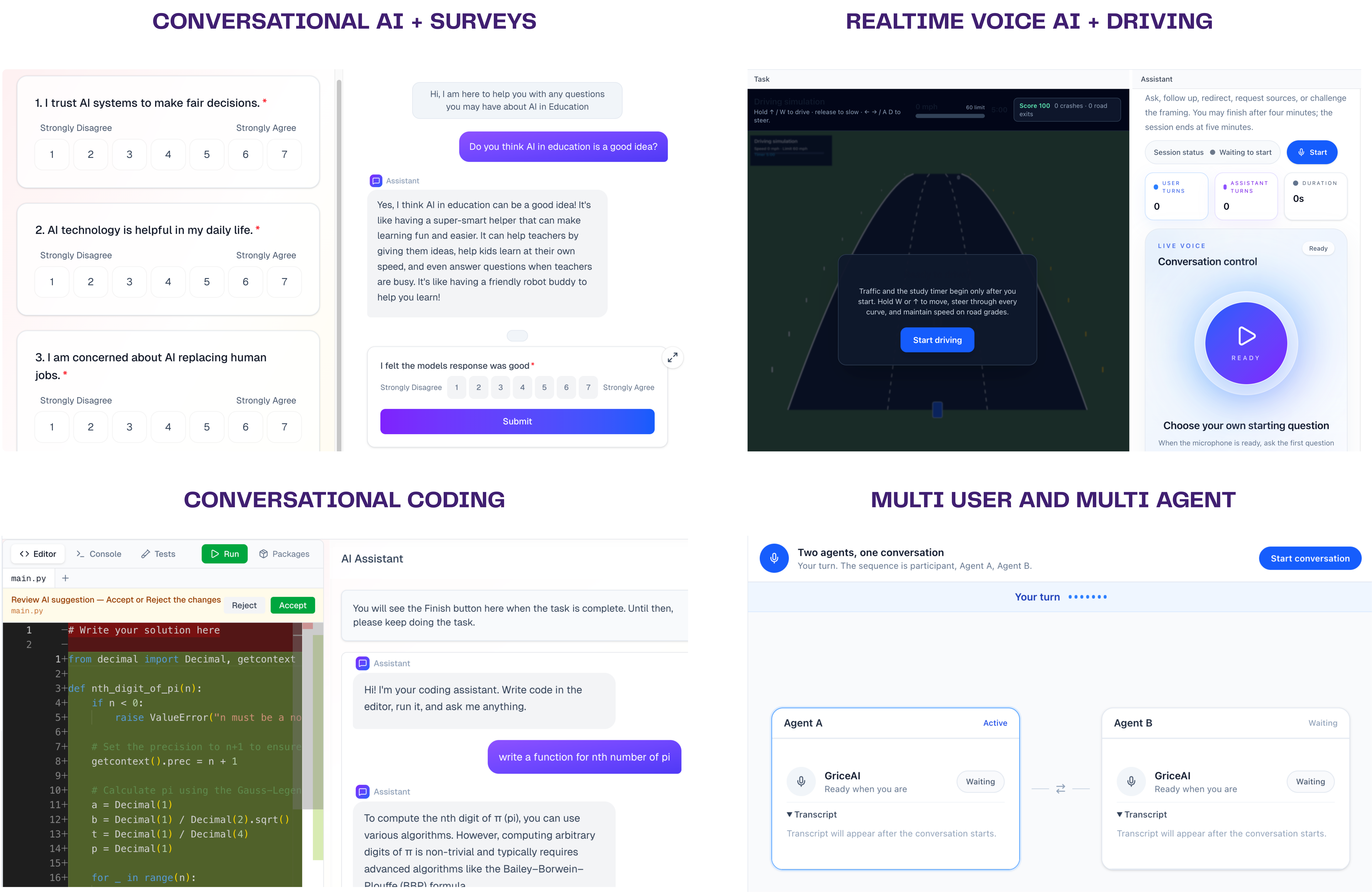}
    \caption{Participant-facing interfaces configured in
    \griceaAccess{Gricea}: conversational AI with integrated surveys
    (top left), real-time voice interaction alongside a driving task
    (top right), conversational coding with an editable code workspace
    (bottom left), and a participant interacting with two conversational
    agents (bottom right).}
    \Description{Four screenshots show participant-facing interfaces.
    The top-left interface combines survey questions, a text
    conversation, and a response-rating item. The top-right interface
    places a driving simulation beside voice interaction controls.
    The bottom-left interface places a code editor with proposed
    changes beside an AI assistant. The bottom-right interface
    displays separate panels for Agent A and Agent B, with indicators
    for participant and agent turns.}
    \label{fig:participant-interfaces}
\end{figure*}
\section{\griceaName: A Platform for Configurable Conversational AI Studies}

\griceaOpen{Gricea} represents controlled conversational AI studies as configurable, executable research artifacts. Its architecture consists of three decoupled layers that operate on this shared representation: a researcher-facing visual authoring environment, a participant-facing execution runtime, and a publication layer for versioning, reuse, and community sharing. Each artifact specifies the experimental procedure and the interactive conditions participants encounter within it. This representation allows complex study conditions to be authored, executed, preserved, and reused within a unified framework rather than reconstructed through ad hoc infrastructure.

\griceaAccess{Gricea} operationalizes study designs as directed graphs. A \textit{Study Flow} graph governs participant progression through the experiment, while nested \textit{Task Flow} graphs specify the behavior of each participant-facing interactive condition. Publication serializes these graphs, their configurations, and their assets into an immutable study version that the runtime engine executes directly.

\subsection{Core System Abstractions}

\griceaOpen{Gricea} represents authored studies through two connected levels of executable graphs. \textit{Study Nodes} form the procedural \textit{Study Flow}, while \textit{Task Nodes} form \textit{Task Flows} embedded in its interactive stages. \textit{Logic Nodes} provide branching, variable assignment, and condition control within these flows. Publication validates the authored definition and makes a specific \textit{Study Version} available as a \textit{Published Study}. \textit{Templates} and \textit{Community Document Collections} provide reusable components and grounding materials, while \textit{Participant Analytics} summarizes the evidence produced during execution.

\paragraph{Study Flow.}
The Study Flow graph encodes the procedural structure through which participants navigate. Its nodes represent participant-facing stages, including instructions, surveys, task entry, annotations, and completion, alongside control operations for branching, randomization, and within-subject ordering. During execution, the engine follows the authored transitions using assignment results and participant-specific state, advancing through control operations until it reaches a stage requiring participant interaction or study completion.

\paragraph{Task Flow.}
The Task Flow graph encodes the runtime behavior of the interactive condition. Its nodes represent operations such as LLM inference calls, retrieval and context injection, real-time voice interaction, loops, participant input, and updates to interface components. The engine executes these operations according to the graph, coordinating model calls and interface changes with participant actions. When the task completes, control returns to the surrounding Study Flow, where information produced during the interaction can inform subsequent stages and measures.

\paragraph{Logic and Condition Control.}
Condition control is represented through graph structure, node configurations, and variables. Branching, variable assignment, weighted randomization, and within-subject routing determine which conditions participants encounter and in what order. Assignments and condition orders are retained in the participant's execution state. Variables connect assigned conditions to task configurations, making explicit how experimental assignment shapes the participant-facing interaction. Researchers can inspect the parameters deliberately varied across conditions alongside the settings held constant within the same authored study.

\subsection{Researcher Authoring Environment}

Researchers author studies visually through no-code, node-based canvas editors for both \textit{Study Flow} and \textit{Task Flow} (Figure~\ref{fig:study-authoring}). The authoring model builds on the staged procedures, condition branches, and task entry points identified in our formative analysis (Section~\ref{sec:formative}), allowing researchers to express experimental logic in the same graphs that drive execution. Researchers can construct the procedure and interactive condition within one environment while retaining separate control over each.

At the study level, researchers add and connect instructions, elicitation stages, branching logic, randomization, within-subject blocks, and task entry points. At the task level, they configure each task's Task Flow and participant interface, including system prompts, retrieved context, model parameters, supported tool settings, interface layout, scaffolds, and modality-specific settings. Study-level variables connect procedural decisions to the task configuration that participants encounter.

The platform separates iterative authoring from participant deployment. Researchers can run a participant-facing preview and revise an editable draft before publication validates and locks the authored graphs, prompts, configuration states, and runtime bindings into a published study version. The participant runtime executes that version while further development proceeds in new drafts, preserving the exact authored condition deployed to participants independently of subsequent changes.

Researchers can inspect assignment and routing in Study Flow, examine their effects on interaction in Task Flow, and preview the participant experience. Publication and execution use this same definition, avoiding a separate translation into bespoke software. Preview complements structural validation by letting researchers check the represented study against their intended design.

\subsection{Participant Runtime and Instrumentation}

The participant runtime combines browser-based rendering with server-side execution of the published study. It resolves the Study Flow for procedural progression, invokes the associated Task Flow upon task entry, and binds participant interactions to the authored graph logic. The browser renders the interface defined for the current stage or task, submits participant actions to the runtime, and applies the resulting updates. The published artifact therefore governs both how participants progress through the study and how the interactive condition responds to their actions.

The runtime supports chat, web-based search tasks, voice interaction, human--AI coding, annotation workflows, image generation, and split-view configurations. These tasks use shared mechanisms for execution, state management, and event collection, allowing researchers to vary the participant-facing experience while preserving a consistent connection between the study configuration and the resulting data.

The instrumentation layer records timestamps, clicks, focus events, scroll depth, input timing, text-edit counts, and task-specific interaction traces according to the study's collection settings. Where enabled, additional capture includes voice recordings, masked browser-session replay, screen recordings, file-upload traces, and eye-tracking data. These records are associated with the participant session and its immutable Study Version, allowing researchers to interpret behavior in relation to the exact configured condition participants encountered.

\subsection{Publication, Reuse, and Community Workflows}

\griceaAccess{Gricea} treats reproducibility as an intrinsic property of the study artifact itself. Each published study is preserved as an immutable version containing its prompts, task parameters, graph structures, assignment logic, and participant-facing materials. These configurations remain inspectable after deployment, and researchers can distribute published studies through shareable links, making the executable study available alongside its written description.

The publication layer supports reuse of both complete studies and their constituent resources. Researchers can publish \textit{Community Templates} containing whole studies or selected nodes and flow segments, share \textit{Community Document Collections} used for RAG grounding, and release de-identified dataset assets. Sharing these resources preserves not only the study's outputs, but also its executable design and the materials used to construct the conversational condition.

A peer researcher can therefore inspect the underlying graph logic and configurations, experience the study live as a participant, and fork the artifact to run an independent replication. The fork provides an editable draft for changing conditions, tasks, or measures without altering the published source. A researcher extending the study can change a selected parameter while retaining the surrounding procedure and materials, then publish and deploy the revised artifact. The source and revised artifacts provide a shared frame of reference for examining what was preserved and what changed, allowing subsequent research to build directly on existing experimental designs.

\subsection{Architecture and Extensibility}

\griceaOpen{Gricea} separates its visual authoring client, execution runtime, and data and publication services so that new capabilities can be added without rebuilding the surrounding study infrastructure. Developers extend task behavior through new node types, each with configuration definitions and an execution handler. Participant-facing components similarly declare the properties researchers can configure and the interaction events they produce. These additions integrate with the existing \textit{Study Flow} and \textit{Task Flow} model.

The architectural objective is to preserve a common study representation as the supported tasks and modalities expand. Extensions for richer multimodal interactions, longitudinal deployments, and community-facing workflows can reuse the same mechanisms for procedural control, publication, and instrumentation, keeping the resulting studies configurable, inspectable, and reusable.

%% file: CHI_2027/04_Evaluation.tex
\section{Evaluation}

\subsection{Evaluation Rationale and Questions}

\griceaAccess{Gricea} is intended to support the replication of existing conversational AI studies and the construction of new studies through a shared, executable representation. We evaluated whether the platform provides the features needed to implement published study designs beyond those included in our formative analysis, and whether researchers from diverse backgrounds can use those features to construct their own studies. Combining replication and usability studies follows prior work on toolkit evaluation, which examines what systems enables and it's usability \citep{ledo2018evaluation, olsen2007evaluating}.

Our evaluation addresses four questions. \textbf{RQ1:} Can \griceaOpen{Gricea} replicate conversational AI study designs beyond those included in our formative analysis? \textbf{RQ2:} Can researchers from diverse backgrounds implement their intended study designs with \griceaOpen{Gricea}? \textbf{RQ3:} Can \griceaAccess{Gricea} represent and execute a diverse range of conversational AI studies? \textbf{RQ4:} What frictions remain in translating research intent into an executable study?

We first conducted a replication study of CUI 2026 papers outside our formative corpus to examine whether \griceaOpen{Gricea} could represent and execute full study configurations or their conversational portions, addressing RQ1 and providing evidence of expressivity for RQ3 (Section~\ref{sec:cui-replication}). We then conducted a usability study in which researchers and AI practitioners authored studies addressing their own research questions, examining their ability to implement intended designs (RQ2), the range of studies they constructed (RQ3), and the frictions they encountered during authoring (RQ4).

\subsection{RQ1: Replication of Published CUI 2026 Studies}
\label{sec:cui-replication}

To evaluate \griceaAccess{Gricea's} support for published conversational AI studies, we examined 37 CUI 2026 full papers, of which 29 reported participant-facing human-subject studies. The full corpus is listed in Appendix~\ref{app:cui-corpus}. These included live conversations, prerecorded conversational stimuli, expert annotation and rating, multimodal interfaces, group interaction, and embodied agents \citep{Bieberstein2026Empathy,Tschopp2026VoiceShopping,Wazzan2026Intent,Hata2026GroupEnvoy,ReyesCruz2026Listening}. The research team extracted study details from the papers and supplementary materials, attempted reconstruction in \griceaOpen{Gricea}, and recorded replication failures.

We assessed replication success by whether \griceaAccess{Gricea} provided the features needed to represent and execute the study procedures, participant-facing interfaces, and conversational task behavior. Complete replication covered the full study configuration, whereas partial replication covered the conversational portions that could be implemented when missing source materials or external dependencies prevented reproduction of the full study. Studies with full replication can still have missing details from studies such as missing participant facing instructions but they do not block representation of the study configurations and CAI tasks. We recorded unavailable information separately from requirements for external hardware or systems to distinguish reporting gaps and scope boundaries from the study features supported by \griceaOpen{Gricea}.

\paragraph{Replication outcome:}
Using \griceaAccess{Gricea}, we successfully replicated full study configurations or their conversational portions from 27 of the 29 eligible papers: 10 completely and 17 partially. \griceaOpen{Gricea} provided the features needed to represent and execute the replicated procedures, interfaces, and conversational tasks, covering prompt-based manipulations, model and stance comparisons, role-play conversations, fixed-media evaluations, repeated sessions, and speech-versus-typing tasks. Missing source materials and external dependencies limited complete replication, while unavailable interview and survey materials prevented replication of the remaining two studies. These results demonstrate that \griceaAccess{Gricea} supports executable study designs beyond those used to inform its representation.

\paragraph{Reporting Gaps and Their Impact on Replication:}
We found incomplete reporting or unavailable materials in 28 of the 29 eligible papers, although not all omissions prevented replication. The most common gaps were questionnaire items, revisions or participant-facing interview wording (21 papers), complete agent prompts, grounding inputs, or configuration rules (15), and stimulus or exercise materials (11), with overlap across categories. When unavailable information was necessary to implement the study configuration, it limited which portions could be replicated.

\paragraph{\griceaOpen{Gricea} External Dependencies and Scope Boundaries} Fourteen papers involved external requirements in at least one phase: robots or wearables, voice-cloning pipelines, specialized applications, or in-person group coordination. These should generally remain external systems connected to \griceaAccess{Gricea} and hence were only partially replicated.

\subsection{Participants}

We recruited $N=10$ participants from diverse disciplinary backgrounds and roles, including researchers, students, faculty, and industry practitioners. The participant summary is provided in the Table \ref{tab:participants}. Of the 10 participants 5 self-identified as Male and 5 self-identified as female. The median age of participants was in the range of 25-34. There were 4 PhDs, 2 Master students, 2 professionals, 1 professor and 1 undergraduate student in our sample. Participants varied in their prior experience with controlled studies and in the kinds of conversational AI questions they had previously explored or hoped to explore. This diversity was intentional since \griceaOpen{Gricea} is meant to onboard researchers from diverse backgrounds by lowering the barriers to conduct user studies.

\subsection{Study Design and Procedure}

Each session was designed to assess both onboarding and open-ended study authoring. The procedure drew on a common pattern in toolkit and platform evaluation: a structured task that helps participants develop the system’s basic mental model, followed by an open-ended task that reveals how they apply the platform to questions that matter to them [8, 98]. Sessions consisted of four phases: pre-task interview, assisted authoring walkthrough, open-ended think-aloud authoring, and post-task reflection. The total study lasted 90 minutes and participants were paid 30\$ through amazon gift cards. Throughout, we recorded the participant screen and audio after obtaining their informed consent.

We began each session with a brief introduction followed by a pre-task interview on their background, prior experience with human subject studies, challenges encountered in running human subject studies, and research questions they were interested in the area of Conversational AI. 

In the assisted authoring walkthrough, participants implemented a fixed research question: \textit{How does the stance of an AI assistant on controversial issues affect users' perceived trust?} During this phase, the facilitator provided guidance on the platform features, while participants retained control of the interface and performed the authoring themselves.

In the open-ended phase, participants were asked to design a conversational AI study of their own choosing using \griceaAccess{Gricea}. They were instructed to think aloud as they worked.

After the authoring tasks, participants completed a post-task survey and took part in a semi-structured interview. These instruments focused on perceived usability, expressivity, reproducibility, likely time savings relative to current workflows, and adoption potential. The interview further probed where participants felt confident, where they felt stuck, and what forms of support would make the platform more useful in their own work. 

\subsection{Analysis Approach}

We analyzed the survey, transcripts and screen recordings using a combination of descriptive summaries and qualitative analysis. Completion, timing, hints, and breakdowns were summarized descriptively across participants. Think-aloud transcripts, facilitator notes, and interview responses were analyzed thematically to identify recurring patterns related to learnability, expressivity, reproducibility, and unmet support needs.

\subsection{Findings}

\subsubsection{Overcoming the Prototyping and Reproducibility Bottleneck}

In the pre-task interviews, participants consistently described current conversational AI workflows as bespoke, costly, and highly unstable. P7 noted, ``There's no standardized way of conducting these studies... these models keep changing over time.'' P6 described building custom interfaces from scratch, noting, ``I manually coded all of that,'' which took ``a few weeks.'' For participants with limited programming experience, this overhead was prohibitive. P9 stated, ``I cannot code so I never thought I would be able to do this on my own, I would have to just pay someone to build it out for me.'' Furthermore, reproducibility emerged as a primary concern. P7 articulated this explicitly: ``If I want to release my system, it is not clear to me how my systems can stand over time and how will my study be replicated and built upon.''

Following the authoring tasks, participants overwhelmingly viewed \griceaOpen{Gricea} as a structural solution to these bottlenecks. P8 estimated that utilizing \griceaAccess{Gricea} would save ``at least 3 to 6 months of effort'' and roughly ``\$25,000 worth of money,'' noting that building their protocol manually ``would have been a nightmare.'' This qualitative enthusiasm was supported by the post-task survey, where participants indicated a strong likelihood to recommend \griceaOpen{Gricea} to colleagues ($M=6.8, SE=0.13$) and expressed high confidence that studies authored in \griceaAccess{Gricea} could be reliably reproduced by other researchers ($M=5.9, SE=0.37$). 

\subsubsection{Participants from diverse backgrounds were able to author studies}

Participants across disciplines successfully mapped their conceptual study designs onto \griceaOpen{Gricea's} dual-flow abstractions, reporting high overall ease of use ($M=6.1, SE=0.31$). Several explicitly stated that the node-based architecture mirrored their internal cognitive models of experimental design. P1 noted that ``all those component building blocks make sense to me'' and could be used ``intuitively'' to build procedures. P3 found that ``the plug, play, click and edit pipeline is really intuitive,'' preserving their ``chain of thought,'' while P4 highlighted that ``the most intuitive part is when you connect those together.'' 

Three platform strengths consistently emerged as critical for supporting researchers from such diverse backgrounds. First, the visual canvas provided necessary architectural clarity; P8 noted, ``The visual interface is quite easy to develop the protocol... The visual overlay made a lot of difference.'' Second, participants valued the unified consolidation of procedure and runtime. P6 appreciated having surveys, AI tasks, and deployment ``contained into one system,'' contrasting it with fragmented legacy workflows. Third, the localized validation and preview mechanisms were highly praised. P7 appreciated ``being able to preview it... and seeing if what I thought is what is actually happening,'' while P3 praised ``the ease with which a study could be validated,'' calling the feature ``super cool.'' These accounts show why inspection matters beyond usability: participants used the representation and preview to connect their intended design to the procedure and interaction the platform would execute.

\begin{table*}[h]
\centering
\small
\renewcommand{\arraystretch}{1.15}
\begin{tabular}{p{1.1cm} p{13.8cm}}
\toprule
\textbf{Participant} & \textbf{Open-ended research question} \\
\midrule
\textbf{P01} & Under what conditions and tasks do users rely on different agentic tool calls, such as web search? \\
\textbf{P02} & How does serendipity differ across web search, generative-AI search, and retrieval-augmented generation? \\
\textbf{P03} & How do proactive versus reactive coding agents shape participants' coding practices? \\
\textbf{P04} & How does AI-assisted creative writing compare to human-assisted creative writing? \\
\textbf{P05} & If people cannot distinguish AI-written text from human-written text, where does the perceived ``AI penalty'' in evaluation come from? \\
\textbf{P06} & How do different personas of AI chatbots affect engineers in mechanical design tasks? \\
\textbf{P07} & How does AI assistance in creativity tasks affect subsequent creativity without AI? \\
\textbf{P08} & How does an LLM-augmented debate with a human moderator affect discussion and decision-making? \\
\textbf{P09} & How do people fact-check information when using generative AI? \\
\textbf{P10} & How do differences in model tier exacerbate the digital divide? \\
\bottomrule
\end{tabular}
\caption{Open-ended research questions participants implemented as valid, runnable \griceaAccess{Gricea} studies. These studies varied across procedures, interfaces, models, surveys, and outcomes, illustrating the range of research designs expressible through the shared study representation.}
\label{tab:participant-rqs}
\end{table*}

\subsubsection{\griceaOpen{Gricea} supported RQs across the Conversational AI Design Space}

The open-ended phase provided robust evidence of \griceaAccess{Gricea's} expressive capacity. Participants successfully generated and operationalized a highly diverse set of research questions, confirming they were able to express the exact study designs they had in mind ($M=5.5, SE=0.40$). 

Implemented studies spanned the breadth of the conversational AI manipulation space. P7 investigated how AI assistance affects subsequent non-AI creativity. P5 explored the ``AI penalty'' by testing evaluations of text with and without AI-disclosure. P2 operationalized a study on serendipity across standard web search, generative search, and RAG architectures. Other participants configured studies on proactive versus reactive coding agents (P3), debates involving two voice agents and a human moderator (P8), conditions prompting users to utilize agentic tool-calls (P1), human fact-checking behaviors with generative AI (P9), and how different tiers of models exacerbate the digital divide (P10). Table~\ref{tab:participant-rqs} lists the research questions.

Participants explicitly confirmed that \griceaOpen{Gricea} easily accommodated these designs. P3 stated, ``All of it was able to be done, and I have been trying to think of studies that I cannot run but I cannot think of one yet.'' P6 reported \griceaAccess{Gricea} supported ``even more than I thought was possible,'' and P8 summarized the platform's capacity as ``all of it and more.'' Even for highly domain-specific protocols—such as P6's mechanical design collaboration or P8's clinical protocols—participants found the platform broadly expressive enough to capture their required experimental manipulations and compliant with medical standards for PII data.

\subsubsection{High Degrees of Freedom and the Need for additional scaffolding}

While \griceaOpen{Gricea’s} flexibility enabled diverse RQs, the high degrees of freedom introduced new methodological frictions. Breakdowns occurred most often for variables and branching: P4 found the randomization logic ``confusing for me for like the branching and the randomized [nodes],'' while P1 struggled with ``how to dynamically insert a prompt'' using variables. 

On a similar note, several participants experienced a conceptual boundary when transitioning from the macro \textit{Study Flow} to the micro \textit{Task Flow}. P6 noted that entering the Task Builder ``brought us to another workflow that was a little confusing.'' P2 found the task canvas overly granular, stating that ``it offers too much detail'' and ``for social scientists who don't need complex configurations they might to abstract away the complexities.'' These frictions suggest that while \griceaAccess{Gricea’s} core abstractions are powerful, researchers, especially those from non-tehcnical background, require stronger scaffolding to manage the flow of variables across nested nodes. P1 and P4 also noted a ``high-learning curve''.

To resolve these issues, participants requested additional support for \emph{methodological mapping}. P7 desired an AI feature to provide ``a basic starting template'' to help structure the flow. P1 articulated an important boundary condition for the platform: \griceaOpen{Gricea} is highly effective when the experimental design is already concrete, but early-stage ideation still requires ``collaborative brainstorming with  AI agents.'' All participants however echoed that once they were used to the platform and ran a few studies they could see themselves getting over these barriers.

%% file: CHI_2027/05_discussion.tex
\section{Discussion}

\subsection{Conversational AI Studies as Research Artifacts}
Studies on conversational AI should remain useful beyond the team that created them, but a publication alone cannot preserve the participant experience when procedures, interfaces, prompts, grounding, and API orchestration remain fragmented across code and configuration. \griceaAccess{Gricea} makes the executable study available for inspection, reproduction, and extension, giving the research community a shared basis for building on an individual study.

Participants valued versioning, inspectability, templates, and shareable links as ways to understand what another researcher built and ran. Preserving the authored condition makes methods easier to audit and interpret, extending the value of these features beyond individual authoring convenience.

Our replication study of CUI 2026 papers revealed a gap between reporting a study for publication and preserving
the information needed to reproduce its implementation (Section 6.2). Although these papers had passed peer review, questionnaire wording, agent prompts and grounding inputs, and stimulus or exercise materials were frequently incomplete or unavailable. When those details were necessary to implement a study, their absence prevented replication of the corresponding portions. A published account can therefore communicate a study's rationale, procedure, and findings while leaving another research team unable to reconstruct the participant experience.

Reporting and reproducibility frameworks call for making methodological choices explicit \citep{guidellm2026,feger2019role,Aguilar_2024}. Our findings motivate preserving executable study artifacts as part of the methodological record alongside the paper, so that the community can inspect and reuse the implementation underlying the reported method. \griceaOpen{Gricea} connects study materials and configurations to the executable study version, preserving them during authoring and execution rather than requiring researchers to reconstruct those connections after publication.
\subsection{Related Research Questions Require Inspectable Study Configurations}

Our CUI reconstructions show that studies can address a shared research question while implementing substantially different experimental conditions. Studies examining uncertainty and reliance implemented that relationship through different tasks and procedures: one compared model and prompting conditions during travel-planning conversations, whereas another manipulated linguistic hedging and decision friction in financial advice with explicit allocation decisions \citep{cui2026_3816227,cui2026_3816231}. Beyond the underlying model, the studies differ in condition assignment, conversational progression, available actions, and the operationalization of reliance. Preserving the configured conditions in a shared representation allows subsequent researchers to inspect an earlier study and identify which parts to retain and which to vary when investigating a related hypothesis.

Studies of conversational contact with an outgroup likewise address related questions through different interaction designs. Prior work has compared individual dialogue with a chatbot expressing a vegan perspective against a static essay, and facilitator-mediated group dialogue with an agent grounded in outgroup members' discussions against a document containing that material \citep{cui2026_3816208,Hata2026GroupEnvoy}. Although both investigate the consequences of interactive exposure to an outgroup perspective, they differ in who participates, whose perspective grounds the agent, how messages are composed, and how attitude change is measured. Preserving the study configurations in a shared artifact allows researchers to compare how related questions have been operationalized and design subsequent studies that retain or vary specific aspects of the interaction.

\subsection{Study Flow and Task Flow as a Methodological Abstraction}

In \griceaAccess{Gricea}, the distinction between \textit{Study Flow} and \textit{Task Flow} clarifies the methodological structure of conversational AI studies as well as organizing the platform. Researchers design procedures that assign participants to conditions, collect pre- and post-task measures, and manage staged progression, while also configuring an interactive runtime in which participants encounter a conversational condition that unfolds turn by turn. The participant experience therefore depends on both the study procedure and the behavior of the interactive task.

Across the papers reviewed in our formative analysis, researchers often communicated study procedures through staged diagrams, branching depictions, and flowchart-like structures. The diagrams externalized how researchers conceptualized and communicated experimental logic, beyond summarizing a paper after the fact. Participants in our evaluation repeatedly described \griceaOpen{Gricea's} visual canvas as matching how they thought about study design, and several highlighted that connecting stages and previewing the resulting flow helped them reason about the experiment more concretely. Visual authoring therefore matters both for usability and for its alignment with how researchers already represent study logic in practice.

The process of developing \griceaAccess{Gricea} suggests that research platforms should organize their abstractions around the choices researchers need to distinguish and control. Study Flow separates assignment, ordering, and measurement from the operations that implement the interactive task, while Task Flow makes those operations inspectable without hiding their dependencies on the surrounding procedure. Separating study procedure from interactive task behavior allows a researcher to change conversational behavior while retaining the study sequence, or to change the experimental design while retaining the task. Because Study Flow and Task Flow also drive execution, the relationships a researcher inspects are the relationships used to implement the study. Block-based authoring makes the methodological choices explicit and reusable, extending work on reusable experimental components and methodological specifications \citep{de_Leeuw_2023,jun2019tea}.

Separating study procedure from interactive task behavior also provides a basis for extending research platforms without fragmenting their methods. A new task component should expose its configurable properties, required inputs, completion conditions, and recorded outputs so that it can participate in the existing procedure and data collection mechanisms. In \griceaOpen{Gricea}, a modality-specific extension can reuse the surrounding study infrastructure while making its methodological consequences visible to the researcher. The component and its connections become part of the executable study representation, preserving the relationship between a configuration change and the participant experience it produces when another research team reuses or extends the study.

The distinction between procedure and interactive task also applies to online studies of decision aids, interactive visualizations, educational interfaces, and collaborative tools, which combine assignment and measurement with a task that responds to participant actions \citep{Chen_Schonger_Wickens_2016,Almaatouq_Becker_Houghton_Paton_Watts_Whiting_2021,Cutler_2026}. The representation could extend to these settings by replacing or adding task components while retaining the surrounding Study Flow and its connection to execution records. Different participant-facing activities could therefore share the same mechanisms for specifying, running, and preserving a study.

\subsection{Lowering Technical Barriers Requires Methodological Scaffolding}

The cost of building controlled conversational AI studies can constrain who conducts them and which questions they explore. Lowering technical barriers broadens participation in producing research, not only in using a platform.

Researchers from different disciplinary backgrounds implemented valid, runnable studies across the research questions in Table~\ref{tab:participant-rqs}. \griceaAccess{Gricea} supported participants' study designs through a common representation while reducing barriers for researchers without systems backgrounds. Sharing the resulting artifacts also allows other teams to inspect and adapt the studies without rebuilding their infrastructure.

Our evaluation also showed where researchers need methodological scaffolding: participants requested help translating research intent into variables, branching, randomization, and validation, and previewing how those choices shape the participant experience. Related systems likewise show the need for guidance alongside greater authoring control \citep{10.1145/3613904.3642016,yao2026throughlens}. In \griceaOpen{Gricea}, methodological scaffolding should help researchers connect a research question to executable study logic while retaining visibility and control over the resulting design.

\subsection{Making Design-to-Implementation Fidelity Inspectable}

With a shared study representation, researchers can examine assignment branches and task configurations, preview the participant experience, and relate execution records to the published Study Version to check design-to-implementation fidelity. P7 described the check as asking whether ``what I thought is what is actually happening''. Inspecting the configured study and its execution lets researchers examine the experimental choices used during execution rather than infer them from a separate implementation.

Even a structurally valid study can diverge from its intended design: a valid within-subject flow may still misrepresent an intended between-subject comparison. Researcher review checks whether the implemented study matches the intended design, while publication preserves the inspected definition and version-linked records expose its execution. Connecting researcher review, the published study definition, and execution records gives the original team and subsequent researchers a common basis for examining fidelity.

\subsection{Open-Science Workflows for Cumulative Conversational-AI Research}

When we scale across a sequence of studies, the community value of \griceaAccess{Gricea} becomes obvious: one team may publish a conversational task and its experimental conditions; another may retain that procedure while changing an agent behavior, interface feature, or participant population to examine a related hypothesis. Preserving the source and revised artifacts makes the methodological relationship between studies inspectable, giving researchers a basis for explaining which findings concern the same configured interaction and which concern a deliberate extension, rather than treating each implementation as an unrelated starting point.

Connected artifacts could help a research community trace evolving hypotheses, identify unresolved comparisons, and synthesize findings across related studies (Figure~\ref{fig:teaser}). Synthesizing findings across studies requires examining differences in populations, measures, and contexts as well as shared configurations. \griceaOpen{Gricea} provides the mechanisms for publishing, inspecting, and reusing artifacts, while the contribution of shared artifacts to cumulative knowledge will depend on sustained community use.

\subsection{AI-Assisted Research Through a Shared Study Representation}

AI-assisted research can build on shared study representations: \citet{liu2026agentnative} argue that agents need executable research artifacts to understand, reproduce, and extend scientific work. In \griceaAccess{Gricea}, authoring agents could propose or modify Study Flow and Task Flow while researchers inspect changes to assignment, conversational behavior, measurement, and the resulting participant experience. Integrating coding assistance with Study Flow and Task Flow would preserve a common method for reviewing and reusing agent-authored studies.

Supporting AI-native research requires distinguishing agents that help conduct research from agents whose behavior is part of the experiment. An authoring agent might retrieve a prior study, propose a controlled variation, or explain a change to its configuration. An agent participating within a study instead operates under the roles, information access, and interaction rules specified by that study, as human--AI research platforms have begun to support \citep{Qian_Tsai_Behr_Hussein_Laugier_Thain_Dixon_2025,yao2026throughlens}. A shared representation can make both authoring changes and within-study agent behavior inspectable, while keeping changes to the research design separate from actions taken within an experimental condition.

Agent-proposed changes could retain a link to the source study and expose which conditions they alter, allowing researchers to review the proposed revisions before accepted changes enter the representation used for deployment and data collection. Human and AI contributions could then be inspected and reused through a common research workflow.

%% file: CHI_2027/06_limitation.tex
\section{Limitations and Future Work}
Gricea is intended for a broad research community. The authoring study demonstrates how researchers from varied backgrounds implemented conversational AI studies, but its sample of ten participants does not capture the full range of prospective researchers and research practices.

The evaluation covers a subspace of the methodologies and configurations that Gricea is intended to support. The reconstructed online components do not capture the facilitated group activities of participatory co-design workshops, situated observations of people using their own devices and assistive technologies, or the physical behavior of embodied agents \citep{cui2026_3816210,cui2026_3816209,ReyesCruz2026Listening}. These methods involve participant activities, physical settings, and researcher involvement beyond the procedures and interactions represented in the reconstructed components.

The published-study evaluation is also limited to CUI 2026. The corpus provides varied conversational interfaces and study procedures, but does not cover the full range of social-science experiments or the complex combinations of task behavior, modalities, and study logic that Gricea permits.

\paragraph{Future work:} Our evaluation highlights three directions for future work. First, to help researchers navigate Gricea's high degrees of freedom, we plan to integrate an AI copilot that proposes initial study configurations from high-level research questions and modifications to existing artifacts. Proposals would remain within the shared study representation and subject to researcher inspection, validation, and preview before publication. Second, addressing feedback that Gricea is currently ``best used when there is [a] clear RQ'' (P1, P10), we will add methodological scaffolding to help researchers brainstorm and identify potential confounders during early-stage ideation. Third, \griceaOpen{Gricea} allows support for running AI participant simulations that could support pre-deployment checks of study paths, interaction behavior and pilot studies, further lowering the cost of conducting human subject studies. \citep{park2023generativeagentsinteractivesimulacra}.

%% file: CHI_2027/appendix.tex
\section{Appendix}
\subsection{Participant Details}
The participant summary is provided in Table \ref{tab:participants}.

\begin{table}[htp!]
\centering
\small
\begin{tabular}{p{0.08\textwidth} p{0.10\textwidth} p{0.10\textwidth} p{0.10\textwidth} p{0.52\textwidth}}
\toprule
\textbf{ID} & \textbf{Role} & \textbf{Conducted studies?} & \textbf{Study count} & \textbf{Primary research interests / intended study} \\
\midrule
P01 & Researcher & Yes & 5--10 & Capability boundary of conversational AI; what tasks can and cannot be automated; human--human / human--agent boundaries. \\
P02 & Master Student & No & N/A & Serendipity in search; comparing web search, GenAI search, and RAG search; detailed search-session logging. \\
P03 & Master Student & No & N.A. & Legal AI and hallucinated precedent; coding agents; mental-health advice and human influence. \\
P04 & PhD & Yes & 1--5 & Social consequences of conversational agents; tutor-like interaction; transfer from human--AI to human--human interaction. \\
P05 & PhD & Yes & 1--5 & AI penalty: how judgments change when people learn a text was AI-generated versus human-generated. \\
P06 & PhD & Yes & 1--5 & Utility of AI systems in design / mechanical collaboration; AI personas in design tasks. \\
P07 & PhD & Yes & 1--5 & Long-term creativity after AI assistance; comparing no-AI, AI-guidance, and answer-like AI support. \\
P08 & Professor & Yes & 10+ & Surgeon knowledge assessment; using LLMs to identify knowledge deficits and support dialogue around experiential expertise. \\
P09 & Professional & Yes & 10+ & How users use physical devices related to healthcare; \\
P10 & Undergraduate Student & No & N/A & N/A \\
\bottomrule
\end{tabular}
\caption{Completed participant records used in the analysis ($n=10$ usable completed records in the attached JSON export). Participants spanned multiple roles and brought diverse conversational-AI study ideas to the authoring task.}
\label{tab:participants}
\end{table}
\input{CHI_2027/cui_2026_corpus}

%% file: CHI_2027/cui_2026_corpus.tex
\subsection{CUI 2026 Replication Corpus}
\label{app:cui-corpus}

Table~\ref{tab:cui-corpus} lists all 29 eligible papers included in the published-study replication evaluation (Section~\ref{sec:cui-replication}).

\begin{table*}[t]
\centering
\small
\renewcommand{\arraystretch}{1.0}
\caption{The complete corpus of 29 eligible CUI 2026 papers that we attempted to recreate.}
\label{tab:cui-corpus}

\begin{tabular}{@{}
  p{\dimexpr.12\linewidth-\tabcolsep\relax}
  >{\raggedright\arraybackslash}
  p{\dimexpr.88\linewidth-\tabcolsep\relax}
@{}}
\toprule
\textbf{Reference} & \textbf{Paper} \\
\midrule
\citep{Bieberstein2026Empathy} & Do Prompt-Level Empathy Instructions Influence User Experience? Evidence From A Controlled Chatbot Study \\
\citep{cui2026_3816199} & AI Echoing in the Backstage: Private AI Consulting May Strengthens Confidence and Limits Depolarization \\
\citep{cui2026_3816227} & Beyond Benchmarks: A User-Centric Framework for Evaluating Large Language Models \\
\citep{cui2026_3816231} & Linguistic Uncertainty Markers for Trust Calibration in AI-Assisted Decision-Making \\
\citep{cui2026_3816203} & Beliefs and Misconceptions around Integrated Conversational AI \\
\citep{cui2026_3816210} & Design and Evaluation of ChatBlend: A Framework and Card Deck for the Human-Centered Design of Mental Health Chatbots in Blended Care \\
\citep{cui2026_3816212} & “More Like a Person’s Voice”: Exploring the Design of Empathetic Virtual Agents for Older Adults \\
\citep{cui2026_3816207} & Beyond Captions: Shaping Imagery Models for Blind or Visually Impaired People During Interacting with AI-powered Conversational Visual Assistant \\
\citep{cui2026_3816209} & “I’M BLIND, ChatGPT”: Interactional Breakdown and Repair in LLM-based Conversational AI for Visually Impaired Users \\
\citep{cui2026_3816226} & LLM-based conversational agents for dyslexia treatment: From chatbots to structured tutors \\
\citep{Wazzan2026Intent} & Comparing Intent Communication Modes for Instruction-based Image Editing \\
\citep{cui2026_3816234} & Butlerliezer: Context- and Receiver-Aware Appropriately Deceptive Auto-Reply System based on Egocentric Video \\
\citep{cui2026_3816223} & Toward Metaphor-Fluid Conversation Design for Voice User Interfaces \\
\citep{cui2026_3816230} & ConvoDojo: Structured LLM-based Sparring Partners for Difficult Workplace Conversations \\
\citep{cui2026_3816218} & AI Interviews the Interviewers: Practitioner Experience and Evaluation of Conversational AI Interviewing \\
\citep{cui2026_3816219} & Sruthi: What Becomes Sayable Through Peer-Mode Reflective Scaffolding in Fieldwork Communication \\
\citep{cui2026_3816213} & SPARC: Exploring Interaction, Sensemaking, and Engagement in AI-Augmented News Reading \\
\citep{cui2026_3816224} & A Comparison of Speech and Typing Input for Creative Generative AI Tasks \\
\citep{cui2026_3816235} & XPLAIN: A Proactive Scaffold Across Speech Processing Stages---Supporting Non-Native Speakers in Real-Time AI-Mediated Turn-Taking \\
\citep{Tschopp2026VoiceShopping} & Does Humanizing Chatbots Promote or Hinder Voice Shopping? \\
\citep{ReyesCruz2026Listening} & Embodied Active Listening: How Non-Verbal Backchannel Behaviour Influences Trust and Perception of a Social Robot \\
\citep{cui2026_3816211} & Exploring Perceptions of Robo-advisors for Personal Investing \\
\citep{Hata2026GroupEnvoy} & GroupEnvoy: A Conversational Agent Speaking for the Outgroup to Foster Intergroup Relations \\
\citep{cui2026_3816206} & Beyond Emotional Mirroring: Understanding Affective Alignment in Artist-in-the-Loop Art Chatbots \\
\citep{cui2026_3816215} & Towards Pedagogy-Grounded Conversational AI Tutors for Interest-Based Learning \\
\citep{cui2026_3816208} & More Interactivity, More Open-Mindedness? The Influence of AI-based Chatbot Interactivity on Attitudes Toward Vegans \\
\citep{cui2026_3816201} & Is ChatGPT Gender-Neutral? Implicit Stereotyping Persists Over Time and With Experience \\
\citep{cui2026_3816200} & When Text-to-Speech Speaks in Your Voice: A Study on Public Perception \\
\citep{cui2026_3816205} & Designing a Feedback Loop Between a Human and Their AI Clones for Science Communication in Museums \\
\bottomrule
\end{tabular}
\end{table*}